\documentclass[preprint, superscriptaddress, showkeys]{revtex4}
 \usepackage[dvips]{graphicx}

\begin{document}

\title{
Exploring the Protein G Helix Free Energy 
Surface by Solute Tempering Metadynamics
}
\date{\today}

\author{Carlo Camilloni}
\affiliation{Department of Physics, University of Milano, via Celoria 16, 20133 Milan, Italy}
\affiliation{INFN, Milan Section, Milan, Italy}
\author{Davide Provasi}
\affiliation{Department of Physics, University of Milano, via Celoria 16, 20133 Milan, Italy}
\affiliation{INFN, Milan Section, Milan, Italy}
\author{Guido Tiana}
\email[Corresponding author: ]{guido.tiana@mi.infn.it}
\altaffiliation[address:] {Department of Physics, via Celoria, 16, 20133 Milano, Italy}
\altaffiliation[phone:]{+39-02-50317221}
\affiliation{Department of Physics, University of Milano, via Celoria 16, 20133 Milan, Italy}
\affiliation{INFN, Milan Section, Milan, Italy}
\author{Ricardo A. Broglia}
\affiliation{Department of Physics, University of Milano, via Celoria 16, 20133 Milan, Italy}
\affiliation{INFN, Milan Section, Milan, Italy}
\affiliation{The Niels Bohr Institute, University of Copenhagen, Blegdamsvej 17, DK-2100 Copenhagen, Denmark}

\thanks{{\bf Short title:} Free energy of protein-G helix}

\begin{abstract}
The free-energy landscape of the $\alpha$-helix of protein G is studied by means of metadynamics coupled 
with a solute tempering algorithm. Metadynamics allows to overcome large energy barriers, whereas solute 
tempering improves the sampling with an affordable computational effort. From the sampled free-energy 
surface we are able to reproduce a number of experimental observations, such as the fact that the lowest 
minimum corresponds to a globular conformation displaying some degree of $\beta$-structure, that the helical 
state is metastable and involves only $65\%$ of the chain. 
The calculations also show that the system populates consistently a $\pi$-helix state 
and that the hydrophobic staple motif is present only in the free-energy minimum associated with the helices,
and contributes to their stabilization. The use of metadynamics coupled 
with solute tempering results then particularly suitable to provide the thermodynamics of a short peptide, 
and its computational efficiency is promising to deal with larger proteins.
\end{abstract}


\maketitle

\section{Introduction}
The B1 domain of streptococcal protein G (Protein G thereafter) is the immunoglobulin binding domain 
of the protein and comprises 56 residues located at its N-terminus. The folding of this small protein has been 
studied thoroughly, and accurate structural and thermodynamic characterizations are available for it. The domain 
is a stable globular folding unit with no disulfide cross-links, and in its native fold a central $\alpha$-helix 
is packed against a four-stranded $\beta$-sheet, formed by two anti-symmetrically disposed $\beta$-hairpins~\cite{gronenborn91}. 
The folding occurs without detectable intermediates. Differential scanning calorimetry~\cite{alexander92a} and 
stopped-flow mixing methods~\cite{alexander92b} show that the protein exhibits a two-state unfolding behavior over a 
wide pH range, and that the kinetics of folding and unfolding can be fit to a single, first-order rate constant 
over a wide temperature range.  

Interestingly, the three fragments of secondary structure have different stabilities when isolated from the 
remainder of the protein; in particular, the second $\beta$-hairpin (comprising residues 41 through 56) is the most 
stable one, while the helix (residue 21 through 40) and the first $\beta$-hairpin (residues 1 through 20) are 
unstructured in water~\cite{blanco95}. Moreover, the $\alpha$-helix fragment has been found to be stabilized by some 
non-native hydrophobic interactions with its C- and N-terminus residues.~\cite{blanco97}. Finally, replacing the helix 
wild-type sequence at residues 21 through 40 with the second-hairpin sequence, the same native fold is obtained, 
suggesting that are the non-local interactions due to the $\beta$-hairpin that determines the fold of protein G~\cite{cregut99}. 

By studying the effects of point mutations in protein G, McCallister, et al.~\cite{McCallister00} suggest that its 
transition state is characterized by a largely structured second $\beta$-hairpin forming three stranded $\beta$-sheet 
with the N-terminal $\beta$-strand of the  first hairpin. In the transition state the helix seems only partially 
formed towards the C-terminus region.

Although such intense experimental work provides interesting information about the folding dynamics of protein G, 
the extremely complex nature of the process makes it hard to rationalize its details, making the problem well suited 
for a computational analysis. While it would be computationally easy to make standard unfolding simulations of a 
20-residues peptide, we pursue a more ambitious goal: to calculate its equilibrium free-energy landscape. 
With standard molecular dynamics simulations, this is reached only after a large number of folding and unfolding 
events, corresponding to many crossing of barriers whose height is much larger than $k_bT$.

To overcome this limit and explore the thermodynamic properties of such systems within all-atom, explicit solvent models, 
several different methods have been proposed. Among them are: a) the parallel tempering method~\cite{hansmann97, sugita99} 
that allows the system to diffuse faster along its phase space by stochastically swapping different replicas with different 
temperature; b) the metadynamics~\cite{laio02, laio05, bussi06a}, where the system is allowed to climb over large 
free-energy barriers by introducing a non-Markovian potential defined as a function of a set of few collective variables 
(CV) which disfavor the exploration of region already sampled. These two different approaches were combined recently~\cite{bussi06} 
to get the free energy surface (FES) of protein-G second $\beta$-hairpin.

The application of such method to larger systems, however, is hampered by the need to use a big number of replicas in 
order to ensure a proper rate of exchange. The acceptance probability of a swap between two replicas is proportional to 
$\exp(\Delta \beta \Delta E) $, where $\Delta \beta$ is the difference of the inverse of replicas temperatures times the 
Boltzmann constant, and $\Delta E$ is the replicas' energy difference. The larger the system (and thus the larger $\Delta E$), 
the smaller the difference in temperature between replicas and thus the larger the number of replicas needed for a fast 
equilibration. This computational limitation prevents the use of parallel tempering for systems larger than $\sim 10$ 
residues. To overcome this limit Liu, et al.~\cite{liu05}, proposed a replica-exchange solute tempering algorithm which reduces the
number of replicas needed for an efficient equilibration, thus allowing the study of larger systems.

In this work we combine metadynamics and solute tempering for the first time, and use it to hike the FES 
of the $\alpha$-helix of the protein G. We show that the combined action of the two approaches constitutes a powerful 
method to study the thermodynamic properties of large complex systems at the all-atom level with explicit solvent. Furthermore
we demonstrate that metadynamics and solute tempering is computationally more affordable than parallel tempering metadynamics.

From the sampled FES we are able to reproduce and to provide an interpretation of several experimental findings.
Our calculations correctly show that the helix is not stable in water, that the metastable helical state is shorter than the 
helix as it is found in the protein. Interestingly also $\pi$-helical structures are found. Furthermore, we observe the presence 
of partial $\beta$-structures in the unfolded region. Moreover we try to characterize some aspects of the helix formation pathway 
and non-native interaction that stabilize the metastable helix.

\section{Materials and Methods}

\subsection{Theoretical framework}

In parallel tempering a set of replicas evolving at increasing temperatures $T_m = (k_B \beta_m)^{-1}$ are swapped within the
replica exchange algorithm, a procedure which improves the correct sampling of the canonical ensemble for all the replicas.

The solute tempering algorithm allows to dramatically reduce the number of replicas needed in a parallel tempering run, by 
allowing the potential energy to scale with temperature in such a way that the molecule of interest appears to get hotter,
but water stays cold as one climbs the replica ladder. Liu, et al.~\cite{liu05} devise a rigorous transformation in which 
acceptance probability for replica exchange, scales only with the number of degrees of freedom of the biomolecule but not with 
the number of water molecules. 

The total interaction energy is written as a sum of three contributions: a protein-protein interaction term $E_{p}$, a 
protein-water interaction $E_{pw}$ and water-water interaction $E_{w}$. The colder replica evolves under the action of the 
true interaction at the physical temperature $T_0 = (k_B \beta_0)^{-1}$ while the warmer replicas, which evolve at $T_m = (k_B \beta_m)^{-1}$,   
have their interaction rescaled as $$ E_m=E_p + \frac{\beta_0}{\beta_m}E_{w}+\sqrt{\frac{\beta_0}{\beta_m}}E_{pw}.$$
Since the potential energy surfaces of all the warm replicas are rescaled, only the cold replica 
sample the correct distribution in the phase space. Following the approach of Bussi, et al.~\cite{bussi06}, 
we complement the improvements due to the replica exchange algorithm 
with those of metadynamics~\cite{laio02, laio05, bussi06a}.

We consider  $N$ replicas of the system. We run them in parallel at increasing temperatures and, with a frequency $\tau_x^{-1}$, 
exchange between two adjacent replicas is attempted. Each replica evolves under the action of the force field and of an
adaptive bias potential $V_m(\vec{s},t)$ calculated using the metadynamics algorithm. This potential acts on a set of collective 
variables $\vec{s}(X)$ defined as a function of the microscopic coordinates $X=\{ \vec{R}_1, \vec{R}_2, \ldots \}$.  
With a given frequency $\tau_G^{-1}$, the bias potential is updated adding a Gaussian hill, so that the potential is given by
$$ V_m(\vec{s}, t) = w_m \sum_{t_k < t}{\exp \big ( - \sum_i \frac{ (s_i-s_i(X_m(t_k)))^2}{2 \sigma_{i}^2} \big ) }, $$ 
where $X_m(t_k)$ are the microscopic coordinates of the $m$-th replica at time $t_k$, $w_m$ gives the height of the hill 
contribution, and $\sigma_i$ controls its width along the $i$-th direction in the CV space.

Since each replica experiences a different potential due to the bias from the non-Markovian metadynamics terms, as well 
as to the scaling from the solute tempering scheme, the acceptance probability must be calculated according to the replica 
exchange prescription with different Hamiltonians~\cite{yan99}. A proposed swap is evaluated according to the Metropolis 
algorithm, accepting the exchange between the $i$-th and the $j$-th replica with probability given by $P = \min(1, \exp(- \Delta ))$, 
where  
\begin{eqnarray*}
\Delta   = &  &  (\beta_i-\beta_j) \big ( E_p(X_j)-E_p(X_i) \big ) \\ 
 & + & (\sqrt{\beta_0 \beta_i}-\sqrt{\beta_0 \beta_j}) \big (E_{pw}(X_j)-E_{pw}(X_i) \big )  \\
 & - & \beta_i \big (V_i(s(X_i))-V_i(s(X_j))  \big ) \\
 & - & \beta_j \big (V_j(s(X_j))-V_j(s(X_i))  \big ).
\end{eqnarray*}
As already remarked, only the colder replica samples the proper FES, that can be extracted from the limit 
for large simulation times of the history-dependent potential~\cite{laio02, laio05, bussi06a}: 
$$ F(\vec{s}) = \lim_{t\rightarrow \infty} V_m(\vec{s},t). $$

To asses the effect of the metadynamics on the convergence properties of the simulation, we also make a control 
simulation where we compared the combined approach just described with a run without metadynamics.  As shown below in detail, 
the approach described in this work combines the improved convergence already reported for the parallel tempering metadynamics 
with a significant reduction in computational effort due to the solute tempering. For example, for a system of
$3\times10^{4}$ atoms only ten replicas are needed to effectively cover a range of more than 300 K, with respect to more than 60 
replicas used for the parallel tempering. 

The metadynamics algorithm, solute tempering and combined solute tempering metadynamics have been implemented by the authors 
on the GROMACS molecular dynamics package~\cite{berendsen95, lindahl01} and are available upon request.

\subsection{Simulation details}

The $\alpha$-helix structure studied corresponds to the fragment from the residues 21 through 40 of protein G (PDB code 1PGB). 
The interactions were described using the GROMOS 53A6 force field~\cite{oostenbrink04, oostenbrink05}, and virtual-site atoms 
for hydrogens were used to speed up the simulation~\cite{miyamoto92, hess97}, allowing the time step for the molecular dynamic 
integration to be as high as 0.004 ps. The system was enclosed in a dodecahedron box of 300 ${\rm nm}^3$ with periodic boundary 
conditions and solvated with 9856 SPCE water molecules~\cite{berendsen87}. The system charge was neutralized adding 2 Na$^+$ ions.
Van der Waals interaction were cut-off at 1.4 nm and the long-range electrostatic interactions were calculated by the particle 
mesh Ewald algorithm~\cite{essman95} with a mesh spaced 0.125~${\rm nm}$. The neighbor list for the non-bonded interactions 
was updated every 5 steps (0.020 ps). The system evolves canonically, thermally coupled with a Nos\'e-Hoover bath~\cite{nose84, hoover85}.

The solvated system was prepared using the following procedure: (1) a steepest descent energy minimization; (2) equilibration 
of the solvent for 100 ps at 100 K, keeping the heavy atoms of the protein constrained with springs of strength 1000 kJ/(nm$^2$ mol); 
followed by another 100 ps at 200 K with the spring constant reduced to 500 kJ/(nm$^2$ mol) and by another 100 ps at 300 K with
springs of 250 kJ/(nm$^2$ mol); (3) a thermal equilibration of the whole system, with a 100 ps dynamics at 300 K, at constant 
volume; (4) a density equilibration with a  100 ps dynamics at 300K and constant pressure, coupling the system to a Berendsen 
barostat, and finally (5) a 2 ns dynamics at 300 K at constant volume to thermalize again the system.

After such preparation, we run 10 replicas of the system (respectively at 300, 325, 352, 381, 413, 447, 484, 524, 568 and 615 K), 
attempting to swap neighbors replicas every 50 steps, $\tau_x = 0.2$~ps. We verify {\it a posteriori} that choosing these 
temperatures we achieve an optimal acceptance rate (between 30\% and 40\%).

Regarding the acceptance rate of the proposed swaps between replicas, we observed a significant increase from the solute tempering 
metadynamics (where it was about 35\%) and the simple solute tempering case, where the same temperature choices reduced the rate to 25\%.

Moreover in ref.~\cite{huang07} Huang, et al. shows that solute tempering is not efficient for proteins because not taking into 
account the water-water interaction energy in the replica exchange acceptance probability reduce greatly the exchanges between 
folded and unfolded conformations. In the particular case under study we have seen that each replica explores the whole temperature range or 
for solute tempering  either for solute tempering metadynamics. However, for the last approach an increase of exchange probability was observed.

The choice of the collective variables included in the metadynamics reflects the physical insight of the problem being addressed.
In order to describe the folding of the helix, we therefore choose as collective variables the radius of gyration of the backbone heavy
atoms and the number hydrogen bonds between backbone atoms of residues 4 sites apart along the sequence ($i$,$i+4$ hydrogen bonds).
Since the algorithm requires the collective variables to be differentiable functions of the microscopic coordinates, we evaluated 
this count as 
$$N_H= \sum_{i=2}^{13}  \frac{1-(r_{i,i+4}/d_0)^n}{1-(r_{i,i+4}/d_0)^m}, $$
with $n=6$ and $m=10$, $d_0=0.32$~nm, and where $r_{i,i+4}$ is the distance between the backbone oxygen of residue $i$ and the 
backbone nitrogen of residue $i+4$.

Clearly, there is a trade-off in the choice of the values of $m$ and $n$. A better discrimination between formed and non-formed H-bonds 
entails that an higher number of conformations display a vanishing value of the gradient $\partial N_H / \partial r_{i,i+4}$.  For 
these conformations, the history-dependent potential has little effect on the dynamics, and thus the algorithm looses its effectiveness 
in correctly sampling all the relevant phase-space.  

Moreover, it is important to notice that the functional form that defines this CV does not contain any angular contribution; thus, 
conformations in which one oxygen at site $i$ falls close to one nitrogen at site $i+4$ will contribute to the CV even if their 
positions do not satisfy the angular criterion determining an H-bond (i.e. $\theta_{\rm H-N-O} < 30^\circ$). Again, including such
criterion would on one hand improve the accuracy with which the CV describes the formation of helices, but on the other hand, 
enlarge the region of the phase space in which the gradient of the CV with respect to the microscopic atomic coordinates vanishes.
The chosen values of the exponents $m$ and $n$ guarantee that the derivatives along the range of distances $r_{i,i+4}$ never
vanish to the numerical precision. This comes with the price of a reduced accuracy in identifying the exact number of formed 
$i-(i+4)$ H-bonds.  

Notice that these variables do not require any specific knowledge of the folded structure. 

We choose a Gaussian height $w_m = 0.7 k_B T_m $, a width $\sigma_{\rm Gyr}$ of 0.01 nm and 
$\sigma_{\rm Hb}$ of 0.1 nm. The bias potential is updated every 125 steps, $\tau_G = 0.5$ ps.

\section{Results and Discussion}

\begin{figure*}[htbp]
\begin{center}
\includegraphics[height=12cm]{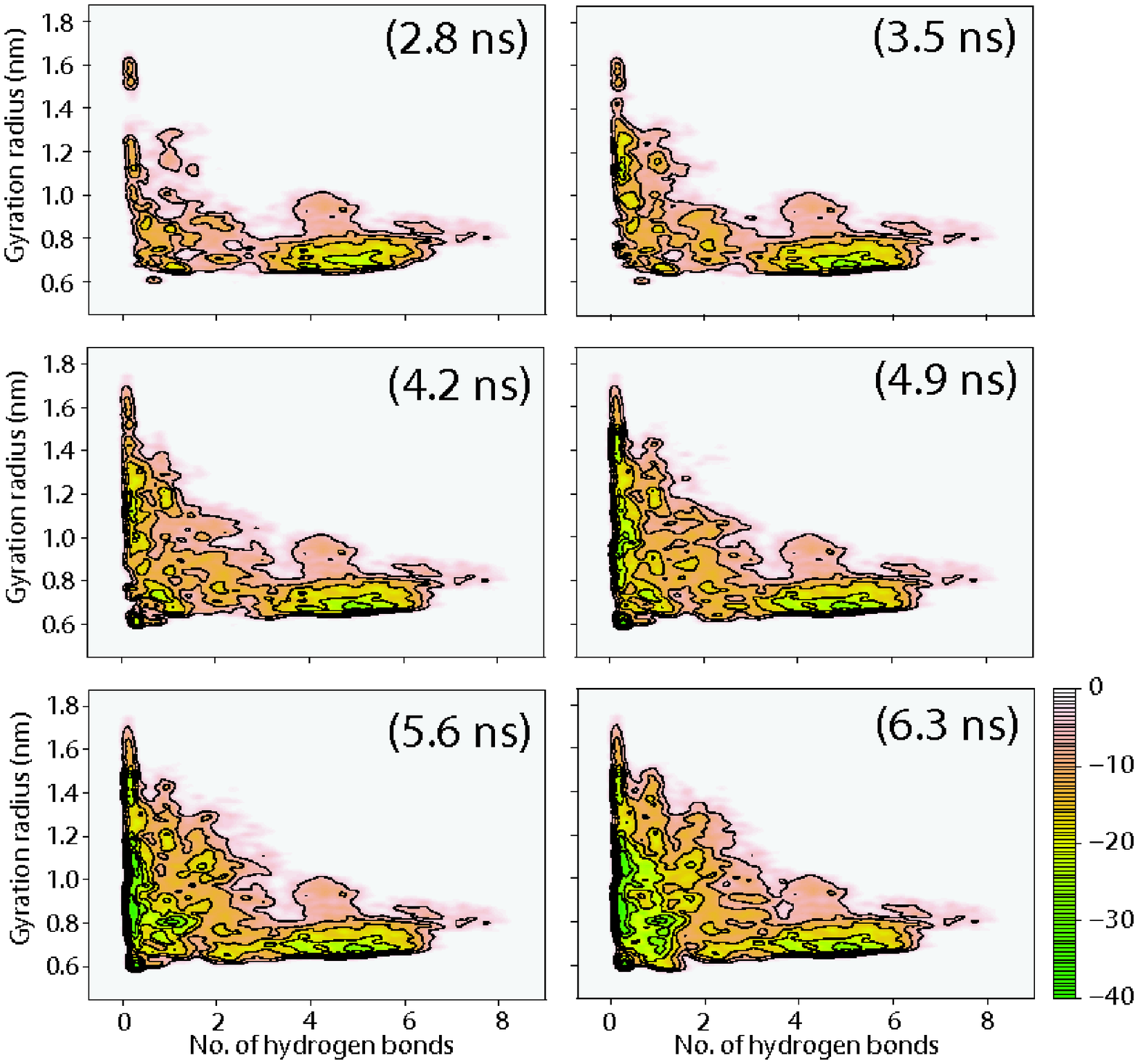}
\caption{
{\bf Evolution and convergence.} Free-energy surface of the helix. The
calculated FES are reported every 0.7 ns until convergence is achieved after
about 6.3 ns. Free-energy is reported in kJ/mol and contour lines are 5 kJ/mol
apart. }
\label{fig1}
\end{center}
\end{figure*}

The free energy of the helix as a function of the CV defined above is shown in Fig.~\ref{fig1}
for different durations of the simulations. The free-energy is depicted every 0.7 ns in order to 
inspect the convergence of the surface along the metadynamics simulation. 
The theoretical boundaries for the CV are from $0$ to $8$ for the hydrogen-bond number and from $0.6$ 
nm to $1.8$ nm for the gyration radius, corresponding to fully compact and fully extended conformations, respectively. 
From these calculations we have found that it takes about 6 ns for metadynamics to explore the whole (accessible range) of the 
phase space. No change in the structure of the minimum and of the transition states
takes place afterward, suggesting that the free energy has converged to its equilibrium shape.

As a further check of equilibration we performed another simulation, applying solute tempering 
metadynamics to a system which, instead of starting from the folded conformation as in the former case,
started from an extended conformation. The resulting FES converged (to the same accuracy) to the
previously calculated FES, within the same time span (data not shown).

\begin{figure*}[htbp]
\begin{center}
\includegraphics[height=12cm]{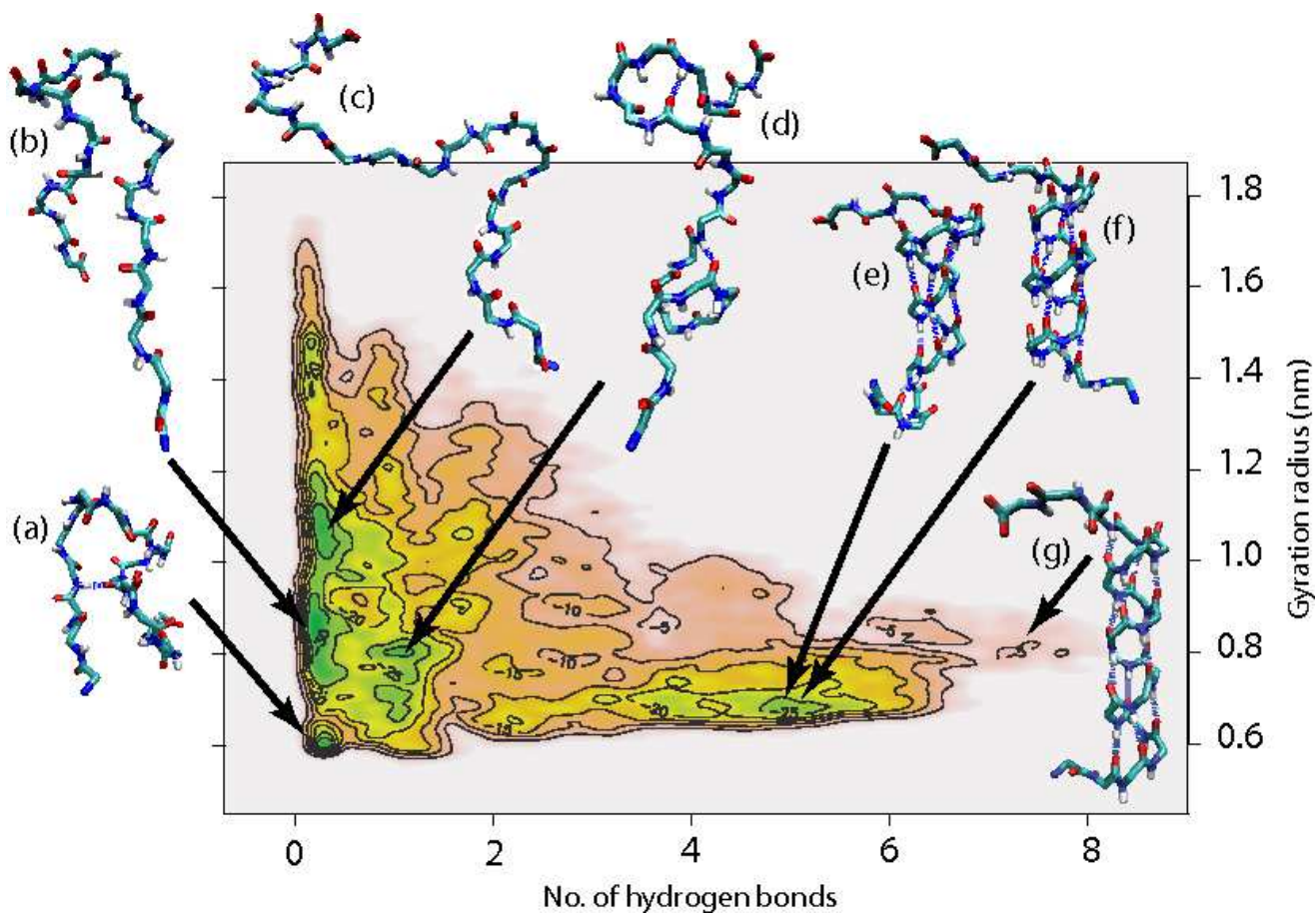}
\caption{
{\bf Relevant meta- and stable structures.}
Converged FES of the helix (after 6.3 ns of metadynamics).
The most representative structures of each local minimum have been represented.
Free-energy is reported in kJ/mol and contour lines are 5 kJ/mol
apart. }
\label{fig2}
\end{center}
\end{figure*}

The most relevant meta-stable structures obtained form the simulations are reported in Fig.~\ref{fig2}, 
along with their position on the FES landscape. In the unfolded region, we observe several 
minima, each of which characterize a different representative structures of the unfolded state.
Referring to the labels indicated in Fig.~\ref{fig2}, state (a) corresponds to a molten globule 
conformation, featuring a very small gyration radius. 
This state is separated from the remainder of the phase space by a high barrier (about 20 kJ/mol).
State (b) corresponds to a $\beta$-bridged structure, with a well defined turn spanning the residues 
28 to 30. Another extended state, corresponding to a random coil conformation is located at an even 
higher gyration radius, and is labeled as state (c). Finally, state (d), displays one $\alpha$-turn 
near the N-terminus region.

As far as the folded region is concerned we find two different conformations in the corresponding 
FES landscape ((e) and (f), Fig. \ref{fig2}), they correspond to an  
$\alpha$-helix (featuring hydrogen bonds between residues $i$ and $i+4$) and a $\pi$-helix (bonds between 
residues $i$ and $i+5$), respectively. Interestingly, the folded helix in the protein (see (g), Fig. \ref{fig2}) displays the last loop in a $i-(i+5)$ 
conformation. The two structures displayed in Fig. \ref{fig2} (e) and (f) are not discriminated by the chosen CV since 
the contribution of the distance between residues $i,i+4$ to the $N_H$ collective variable in the $\pi$-helix is still large. 
Consequently, we cannot assess, even qualitatively, the relative stability of the two structures.

Finally, notice that the structure of the helix in the protein, indicated in Fig.~\ref{fig2} as (g), lies at a even 
higher value of the hydrogen-bond collective variable than do the conformations (e) and (f) (Fig. \ref{fig2}), and does not correspond to a minimum.

\begin{figure}[htbp]
\begin{center}
\includegraphics[height=6cm]{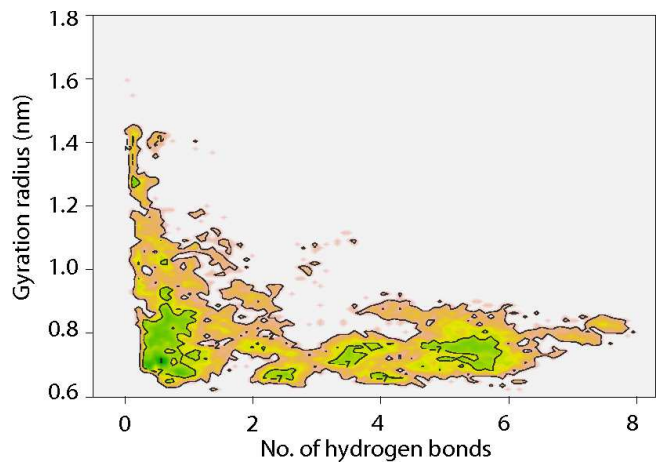}
\caption{
{\bf Solute tempering simulation.}
Free-energy surface of the helix extracted from a 8 ns solute tempering simulation.
Despite the longer sampling, a comparison with the solute tempering
metadynamics results (cfr. Fig.~\ref{fig2}) shows that not all the relevant
configurations have been reached.}
\label{fig3}
\end{center}
\end{figure}

In~\cite{bussi06} it was shown that the combination of parallel tempering and metadynamics shows a significant
improvement in convergence with respect to the simple parallel tempering case. We thus expect that a similar 
synergy be present also in the case under discussion, since solute tempering is expected to behave much like parallel tempering.
To check that this is indeed the case, we performed a solute tempering simulation for 8 ns on the system, monitoring
the behavior of the two CV every 125 steps, (0.5 ps). A FES was extracted from 
such run and is shown in Fig.~\ref{fig3}.

As far as the solute tempering simulation is concerned, it is clear from Fig.~\ref{fig3} that neither the folded nor 
the unfolded region is converged even after 8 ns, and that the area of the phase space explored is significantly smaller than in the
case reported in Fig.~\ref{fig2}.  
Another indication of the broader exploration of the configuration space carried out by  solute tempering metadynamics 
compared to simple solute tempering, is given by the number of clusters explored. 
Clustering the structures in the trajectory with a linkage method (using a cutoff of 0.15 nm, and considering the backbone
RMSD), we obtain 163 clusters for the first 6.3 ns of the solute tempering trajectory, and 398 clusters for the solute 
tempering metadynamics. Even extending the solute tempering to 8 ns, only 248 clusters are sampled.  

Experimental investigations carried out either through nuclear magnetic resonance or circular dichroism~\cite{blanco95, blanco97} 
have been done on a peptide identical to the one studied in this work but for the V21G mutation.  
The helical structure is found not to be stable in water; in particular CD spectroscopy allows to estimate a possible 
helical content of only 5\% at 270 K. This value is in qualitative agreement with the estimate which 
can be done with the help of a two-state picture of the calculated free-energy at 300 K.
In fact, we obtain a free-energy difference between the folded and the unfolded states of $\Delta G_{FU} \approx 9$~kJ/mol, giving 
2\% for the probability of being in an helical state. Furthermore, the same experimental work indicate that the 
helical region in water comprises the residues from D22 to Q32 in accordance with the observed structures in (e), at 
variance with the longer range in the protein structure (g) that extends up to N37.
We notice that as far as the $\pi$-helix (f) is concerned, our results show that these structures 
are helical up to N37 even in water.
 
Another feature of the calculated FES which can be compared with experimental evidence is the presence of partial $\beta$-structures.
Nuclear magnetic resonance data~\cite{blanco95} shows that residues K28 to K31 are significantly populating 
the $\beta$-region of the ($\phi$, $\psi$) space. This is consistent with the finding of very stable conformations 
within such region of the Ramachandran plot in configurations (b).

\begin{figure}
\begin{center}
\includegraphics[height=6cm]{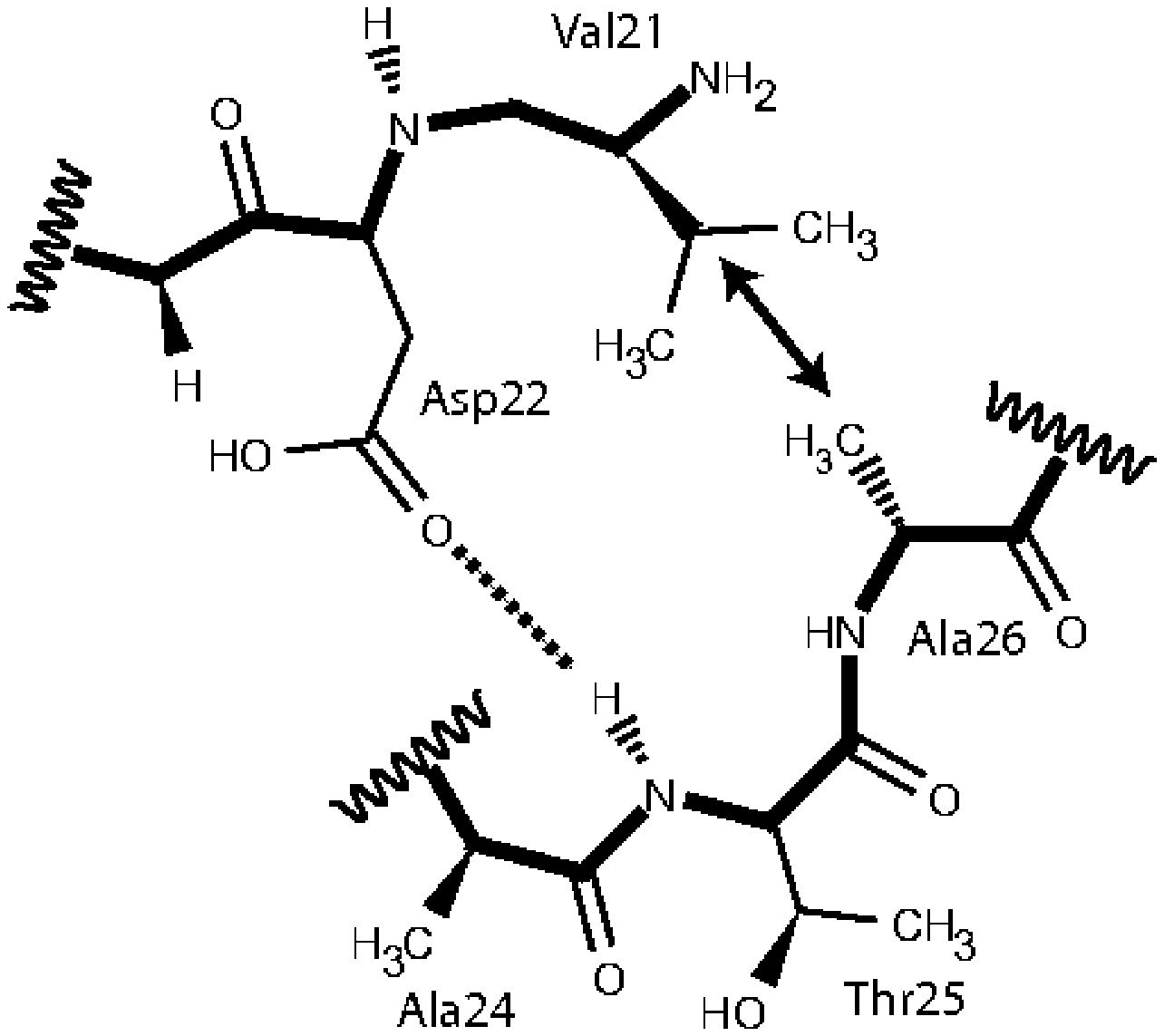}
\caption{
{\bf Hydrophobic staple motif.} Schematic representation of the hydrophobic
staple motif observed in the folded conformations (structures (e) and (f) in
Fig.~\ref{fig3}). The observed contact between V21 and A26 is indicated by a
two-way arrow, while the observed hydrogen bond between D22 and T25 with a
dotted line. } 
\label{fig4}
\end{center}
\end{figure}

An interesting point concerning the stability of the helix regards the presence of the so-called hydrophobic 
staple motif, Fig.~\ref{fig4}. Such motif is not observed in the protein structure, despite the favorable sequence 
at the N-terminus of the helix. Blanco et al.~\cite{blanco97} however, report nuclear magnetic resonance results which 
indicate the likely presence of such stabilizing motif in the isolated helix in water. 

Analyzing the most representative structures in the folded conformations (structures $(e)$ and $(f)$ in Fig.~\ref{fig2}), 
we checked that the side-chain of V21 and A26 are in contact, as well as those of D22 and T25, indicating that indeed the 
staple motif stabilized both the $\alpha$- and the $\pi$- helical structures.

A further analysis which is usually done starting from FES consists in the identification of the most 
probable kinetic trajectory identified with the minimum pathway between pairs of thermodynamic states~\cite{bolhuis02, bonomi07}. 
For this purpose, a careful choice of the CV is critical~\cite{bolhuis02}. On the contrary, the use of the solute 
tempering coupled with metadynamics used in the present work is more tolerant to a loose choice of the CV. An example 
of this issue is provided by the states labeled (e) and (f) in Fig.~\ref{fig2}. This is a single minimum in the free energy 
calculated as a function of $N_H$ and $R_g$, but it contains two sets of conformations, that is $\alpha$- and $\pi$-helices, 
which are certainly separated by energy barriers. In other words, $\alpha$- and $\pi$-helices would require, 
to be distinguished in solute tempering metadynamics simulation, more (or different) CV. From a kinetic point of view, 
this is a serious obstacle for a correct description 
of the folding trajectory. On the other hand, the use of these CV allows solute tempering metadynamics for an efficient 
reconstruction of the free energy landscape.

What can be done instead is to describe the formation of the helix only from a qualitative point of view,
analyzing the features of a representative set of conformations extracted from the local-minimum basin
(i.e. the extended folded region around the points (e) and (f) in Fig.~\ref{fig2}). 
We define operatively the basin as the set of conformations displaying a number of H-bonds between 3 and 
5 and a gyration radius between 0.70 nm and 0.72 nm. 
One can observe that the structures with fewer H-bonds comprise two-formed 
loops in the region from 25 to 33, i.e. in the N-terminus half of the chain. 
No structures with formed loops in the C-terminus are observed. 
One can thus infer that, once these loops are formed, the helix grows on both sides, reaching (e) and (f).
This is consistent with the fact that the primary sequence of the peptide displays a larger helical propensity 
in the N-terminus region than in the C-terminus region~\cite{blanco97}. 
This result seems to point against the suggestion reported in~\cite{McCallister00}, where Baker and coworkers
suggest, on the basis of side-directed mutagenesis studies on the entire protein, that the helix grows from the 
C-terminal.
As far as the staple motif is concerned, an analysis of the same conformations shows that the contact between 
V21 and A26 and the hydrogen bond between D22 and T25 are formed only for structures where the helix is folded 
completely, i.e. in (e) and (f).  No presence of the staple motif is observed when the helix displays only part 
of the turns formed.

\section{Conclusions}

We have calculated and analyzed the FES of a fragment of Protein G corresponding to a 
helical structure in the protein, efficiently sampling the conformations of the system using a combined solute 
tempering and metadynamics algorithm.

We show that the proposed algorithm is able to explore a broad region of the phase space and allows to accurately estimate 
the associated free energy within affordable computational effort.
Several experimental findings can be reproduced and explained with help of the simulation. In particular, we find that 
the helix is not stable in water, and that the metastable $\alpha$-helical state is shorter than the helix found in 
the native state of the protein. We also observe the presence of partial $\beta$-structures in the unfolded state.
Furthermore, the simulations indicates that the peptide populates not only an $\alpha$-helical conformation, but also
a $\pi$-helix, a result which can be tested experimentally. 

As far as the dynamics of helix formation is concerned, the simulation indicate that both kinds of helices grow from the 
N-terminus half. The role of non-native interactions in stabilizing the helical fold is also confirmed by the presence of an 
hydrophobic staple motif, also located at the N-terminus of the helix.

Solute tempering metadynamics shows remarkable robustness against a partial knowledge of all the relevant
slow degrees-of-freedom characterizing the systems, guaranteeing an exhaustive free-energy sampling in spite of
the use of degenerate collective variables (which, for example, are not able to distinguish between $\alpha$- and
$\pi$-helices).

Summing up, the thermodynamics of two of the most important secondary structures of protein G have been characterized by 
two different flavors of replica exchange metadynamics methods (the $\alpha$-helix by us and the second $\beta$-
hairpin by Bussi, at al.~\cite{bussi06}).
The efficiency of the solute tempering method described here 
and its speed, makes it particularly suited to attack larger systems, in particular the whole of protein G.

\clearpage



\clearpage
\begin{thebibliography}{99}

\bibitem{gronenborn91} Gronenborn AM, Filpula DR, Essig NZ, Achari A, Whitlow M, Wingfield PT, Clore GM. 
{\it A novel, highly stable fold of the immunoglobulin binding domain of streptococcal protein G}. Science 1991; 253:657--661.

\bibitem{alexander92a} Alexander P, Orban J, Bryan P. {\it Kinetic Analysis of Folding and Unfolding the 56 Amino Acid IgG-Binding
Domain of Streptococcal Protein G}. Biochem. 1992; 31:7243--7248.

\bibitem{alexander92b} Alexander P, Fahnestock S, Lee T, Orban J, Bryan P. 
{\it Thermodynamic Analysis of the Folding of the Streptococcal Protein G
IgG-Binding Domains B1 and B2: Why Small Proteins Tend To Have High
Denaturation Temperatures}. Biochem. 1992; 31:3597--3603.

\bibitem{blanco95} Blanco FJ, Serrano L. {\it Folding of protein G B1 domain studied by the
 conformation characterization of fragments
comprising its secondary structure elements}. Eur. J. Biochem. 1995; 230:634--649.

\bibitem{blanco97} Blanco FJ, Ortiz AR, Serrano L. {\it Role of nonnative interaction in the folding 
of the protein G B1 domain as inferred from the conformational analysis of the $\alpha$-helix fragment}. Fold. Des. 1997; 2:123--133.

\bibitem{cregut99} Cregut D, Civera C, Macias MJ, Wallon G, Serrano L. {\it A tale of two secondary structure elements: when
a $\beta$-hairpin becomes an $\alpha$-helix}. J. Mol. Biol. 1999; 292:389--401.

\bibitem{McCallister00} McCallister EL, Alm E, Baker D. {\it Critical role of $\beta$-hairpin formation in protein G folding}. 
Nat. Struct. Biol. 2000; 7:669--673.

\bibitem{hansmann97} Hansmann UHE. {\it Parallel Tempering Algorithm for Conformational Studies of Biological Molecules}.
Chem. Phys. Lett. 1997; 281:140--150.

\bibitem{sugita99} Sugita Y, Okamoto Y. {\it Replica-exchange molecular dynamics method for protein folding}. 
Chem. Phys. Lett. 1999; 314:141--151.

\bibitem{laio02} Laio A, Parrinello M. {\it Escaping free-energy minima}. Proc. Natl. Acad. Sci. U.S.A. 2002; 99:12562--12566.

\bibitem{laio05} Laio A, Rodriguez-Fortea A, Gervasio FL, Ceccarelli M, Parrinello M. 
{\it Assessing the accuracy of metadynamics}. J. Phys. Chem. B 2005; 109:6714--6721.

\bibitem{bussi06a} Bussi G, Laio A, Parrinello M. {\it Equilibrium free energies from nonequilibrium metadynamics}.
Phys. Rev. Lett. 2006; 96:090601.

\bibitem{bussi06} Bussi G, Gervasio FL, Laio A, Parrinello M. {\it Free-Energy Landscape for $\beta$ Hairpin Folding from
Combined Parallel Tempering and Metadynamics}. J. Am. Chem. Soc. 2006; 128:13435--13441.

\bibitem{liu05} Liu P, Kim B, Friesner RA, Berne BJ. {\it Replica exchange with solute tempering: A method for sampling
biological systems in explicit water}. Proc. Natl. Acad. Sci. U.S.A. 2005; 102:13749--13654.

\bibitem{yan99} Yan Q, de Pablo JJ. {\it Hyper-parallel tempering Monte Carlo: Application to the Lennard-Jones fluid 
and the restricted primitive model}. J. Chem. Phys. 1999; 111:9509--9516.

\bibitem{berendsen95} Berendsen HJC, van der Spoel D, van Drunen R. {\it GROMACS: A
message-passing parallel molecular dynamics implementation}. Comp. Phys. Comm. 1995; 91:43--56.

\bibitem{lindahl01} Lindahl E, Hess B, van der Spoel D. {\it GROMACS 3.0: A package
for molecular simulation and trajectory analysis}. J. Mol. Mod. 2001; 7:306--317.

\bibitem{oostenbrink04} Oostenbrink C, Villa A, Mark AE, Van Gunsteren WF. {\it A biomolecular force field based on the free enthalpy of
hydration and solvation: The GROMOS force-field parameter sets 53A5 and 53A6}. J. Comp. Chem. 2004; 25:1656--1676.

\bibitem{oostenbrink05} Oostenbrink C, Soares TA, van der Vegt NFA, van Gunsteren WF. {\it Validation of the 53A6 GROMOS force field}.
Eur. Biophys. J. 2005; 34:273--284.

\bibitem {miyamoto92} Miyamoto S, Kollman PA.
{\it SETTLE: An Analytical Version of the SHAKE and RATTLE Algorithms for Rigid Water Models}.
J. Comp. Chem. 1992; 13:952--962.

\bibitem{hess97} Hess B, Bekker H, Berendsen HJC, Fraaije JGEM.
{\it LINCS: A Linear Constraint Solver for molecular simulations}. 
J. Comp. Chem. 1997; 18:1463--1472.

\bibitem {berendsen87} Berendsen HJC, Grigera JR, Straatsma TP. {\it The Missing Term in Effective Pair Potentials}.
J. Phys. Chem. 1987; 91:6269--6271.

\bibitem {essman95} Essman U, Perela L, Berkowitz ML, Darden T, Lee H, Pedersen LG.
{\it A smooth particle mesh Ewald method}. J. Chem. Phys. 1995; 103:8577--8592.

\bibitem{nose84} Nos\'e S. {\it A molecular dynamics method for simulations in the canonical ensemble}. Mol. Phys. 1984; 52:255--268.

\bibitem{hoover85} Hoover WG. {\it Canonical dynamics: equilibrium phase-space distributions}. 
Phys. Rev. A 1985; 31:1695--1697.

\bibitem{huang07} Huang X, Hagen M, Kim B, Friesner RA, Zhou R, Berne BJ.
{\it Replica exchange with solute tempering: efficiency in large scale systems}. J. Phys. Chem. B 2007; 111:5405--5410.

\bibitem{bolhuis02} Bolhuis PG, Chandler D, Dellago C, Geissler PL. {\it Transition path sampling: throwing ropes over
rough mountain passes, in the dark}. Annu. Rev. Phys. Chem. 2002; 53:291--318.

\bibitem{bonomi07} Bonomi M, Gervasio FL, Tiana G, Provasi D, Broglia RA, Parrinello M. 
{\it Mechanistic insight in the folding inhibition of the HIV--1 Protease by a small peptide}.
Biophys. J. 2007; (in press).

\end{thebibliography}
\end{document}